\documentclass[% reprint,
preprint,
 amsmath,amssymb,aps,prb,]{revtex4-1}

\usepackage{graphicx}% Include figure files
\usepackage{dcolumn}% Align table columns on decimal point
\usepackage{bm}% bold math
\begin{document}

\preprint{APS/123-QED}

\title{Cage-related crystal-field effects in the paramagnetic range of PrB$_6$}% Force line breaks with \\

\author{Mehdi Amara}
\email{mehdi.amara@neel.cnrs.fr}
\author{Christine Opagiste}
\affiliation{Univ. Grenoble Alpes, CNRS, Grenoble INP\footnote{Institute of Engineering Univ. Grenoble Alpes.}, Institut N\'eel, 38000 Grenoble, France}
\author{Natalya Yu. Shitsevalova}
\affiliation{Institute for Problems of Materials Science, National Academy of Sciences of Ukraine, Krzhyzhanovsky str.3, 03142 Kiev, Ukraine}%

\date{\today}% It is always \today, today,

\begin{abstract}
New insights into the paramagnetic range of PrB$_6$ below 50 K are revealed through specific-heat, thermal-expansion and magnetic measurements. The latent heats of the two successive magnetic transitions, together with an improved description of the lattice contribution, enable a detailed determination of the thermal variation of the magnetic entropy. Immediately above the first-order magnetic ordering temperature ($T_N = 6.7$ K), the magnetic entropy per Pr$^{3+}$ is close to that of a doublet, instead of the triplet crystal-field ground state identified by inelastic neutron scattering. Temperatures in excess of 30 K are required to reach the entropy of the threefold-degenerate ground state. This is reminiscent of the CeB$_6$ case and, similarly, strongly suggests a dynamical splitting of the crystal-field ground state within the boron cage. Further evidence of this splitting is provided by the analysis of the low-temperature paramagnetic susceptibility, which resembles that of a non-magnetic ground state. In addition, a large thermal-expansion anomaly is observed below 20 K, consistent with the expected centrifugal Jahn-Teller effect.
\end{abstract}

\maketitle

%\tableofcontents

\section{Introduction}
One of the primary influences on a rare-earth ion in a crystal results from its electrostatic environment, which is accounted for by the Crystalline Electric Field (CEF). To quantitatively interpret the magnetic properties of a rare-earth compound, a mandatory step is the determination of its CEF energy scheme(s). This defines the energy levels and wave-functions that, under external perturbations, temperature variations, or internal couplings, determine the magnetic phenomena. One of the early successes of group theory applied to quantum mechanics is the treatment of the CEF problem\cite{bethe1929}. Once the point-group symmetry of the rare-earth site has been established from crystallographic analysis, group theory allows to unambiguously define how the Russell-Saunders ground-state multiplet of the rare-earth ion is split under the influence of the CEF. Energy levels, wave-functions and degeneracies are precisely defined, although their relative ordering and energy separations remain to be determined, usually from spectroscopic techniques.

However, the picture of a perfectly symmetric environment for the rare-earth is difficult to reconcile with the principles of quantum mechanics. Even at zero temperature, quantum fluctuations should blur this idealised picture upon which the group-theoretical approach relies. Although mathematically rigorous, the method relies on an approximation, which is to neglect the influence of fluctuations in the atomic positions. Under conditions that favour such fluctuations, the limits of this approximation may become apparent. One such context is that of rare-earth cage compounds, in which a lanthanide ion occupies an oversized cage, allowing for large spatial fluctuations of the magnetic ion. In the case of a cage center of high symmetry, CEF levels with a degeneracy exceeding the Kramers minimum are favoured. As we previously pointed out\cite{Amara2019}, this high degeneracy cannot survive a displacement from the center; the energy level is split and the rare-earth movement results in a significant energy broadening. If the degeneracy $N_{GS}$ of the crystal-field ground state at the center of the cage exceeds the Kramers minimum, a specific energy scale, associated with a temperature $T_{GS}$, is introduced that is influential already in the paramagnetic range. When cooling the sample below $T_{GS}$, specific signatures are expected, in particular: 
\begin{itemize}
\item[-] a progressive reduction in the magnetic entropy, which falls below the $R \ln N_{GS}$ value,  
\item[-] diminished ionic susceptibilities, particularly the magnetic ones, due to the reduced degeneracy,
\item[-] a thermal expansion anomaly that reflects the centrifugal Jahn-Teller effect. An off-center positions of the rare-earth ion can lower its electrostatic energy, leading to an expansion of its spatial distribution. This affects the system's volume but, unlike a cooperative Jahn-Teller effect\cite{Kanamori1960}, without an overall lattice symmetry lowering.
\end{itemize}

Among the most investigated rare-earth cage systems are the elements from the rare-earth hexaborides series (RB$_6$). These compounds crystallize in the CaB$_6$-type structure, with the rare-earth ions forming a simple cubic lattice intercalated with that of the boron octahedra\cite{Stackelberg1932}. The RB$_6$ compounds are, for the most part, metallic, allowing significant exchange interactions. Yet, the magnetic properties of these compounds often depart, even in the paramagnetic range, from the descriptions that account for the influence of the CEF and exchange interactions. Prof. T. Kasuya\cite{KasuyaNMR97, Kasu1997} first highlighted a feature long overlooked: the RB$_6$ compounds form a cubic lattice of oversized boron cages, within which the rare-earth ion possesses a rather large freedom of movement. Beyond considerations based on the geometry and ionic radii, the latitude of movement of the rare-earth is attested by the phonon dispersion curves: the two lowest branches can be readily interpreted as resulting from the anti-crossing between the flat “rattling” mode of the rare-earth ion and the acoustic branch of the boron framework\cite{Amara2019}.

Further support for the cage interpretation arises from the analysis of the antiferromagnetism in this series. At low temperature, all elements with trivalent magnetic rare-earth ions are antiferromagnets. In most instances, the magnetic ordering is first-order, with an unusual $\left\langle \frac{1}{4}\;\frac{1}{4}\;\frac{1}{2} \right\rangle$ wave vector. In GdB$_6$, TbB$_6$, and PrB$_6$, X-ray diffraction experiments show that, alongside the antiferromagnetic order, static displacement waves of the rare-earth guest stabilize\cite{Amara2005, Amara2010, Walker2009}, with amplitudes, relative to the lattice parameter, approaching 1\%. This indicates that the displacement of the rare-earth ions plays an active role in minimizing the exchange energy\cite{Amara2005}. The exchange coupling correction, associated with a displacement wave, is shown to be energetically favorable for only a few, very specific, magnetic wave vectors, notably $\left\langle \frac{1}{4}\;\frac{1}{4}\;\frac{1}{2} \right\rangle$, which are recurrent in the RB$_6$ series. These displacements, whose emergence reinforces the exchange coupling at $T_N$, provide a straightforward explanation for the first-order nature of the magnetic transition.

In RB$_6$ compounds, the point-group symmetry at the rare-earth site, at the centre of the cage, is the octahedral $O_h$ group. The analysis of inelastic neutron scattering experiments has provided the CEF schemes for the light rare-earth elements CeB$_6$\cite{Zirngiebl1984, Loewenhaupt1985}, PrB$_6$\cite{Loewenhaupt1986} and NdB$_6$\cite{Loewenhaupt1986}. These CEF schemes have in common unusually large energy separations, defining well-isolated ground-state levels. In all three compounds the CEF ground state is non-Kramers: a $\Gamma_8$ quartet for CeB$_6$, a $\Gamma_5$ triplet for PrB$_6$ and a $\Gamma_8$ quartet for NdB$_6$. Exposed to a low-symmetry CEF, the quartets will split into two doublets and the triplet into three singlets. These three elements from the hexaboride series are then promising candidates when looking for cage CEF related anomalies such as those listed above.

In the paramagnetic range of CeB$_6$, an improved description of the phonons contributions to the specific heat has allowed for a better definition of the magnetic entropy thermal variation in the paramagnetic range\cite{Amara2020}. Below 30 K, the magnetic entropy drops below the CEF ground state quartet value, approaching the value for a doublet at the 3.3 K ordering temperature. In the same temperature range, neutron spectroscopy investigations\cite{Zirngiebl1984}, show an increase in the energy separation between the $\Gamma _8$ ground state and the excited $\Gamma _7$, another likely signature of a broadened ground state. A model accounting for a distributed crystal field inside the cage provides a quantitative description for these anomalies\cite{Amara2020}. As regards the expected thermal expansion anomaly, it is indeed reported in the work by Novikov et al.\cite{Novikov2017} as a large increase in the linear expansion coefficient below 20 K, which is confirmed by our own (unpublished) measurements.

In PrB$_6$, the CEF ground state is a $\Gamma_5$ triplet, well below (314 K) the first excited $\Gamma_3$ doublet. Unlike in CeB$_6$, no temperature-dependent shift of the neutron scattering lines could be resolved in the experiments of Ref.~\onlinecite{Loewenhaupt1986}. However, more recent inelastic neutron scattering experiments\cite{Lazukov2007, Lazukov2016} show that, at $T = 20$ K, well above the magnetic transition temperature, the quasi-elastic peak widens significantly (to widths exceeding 2 meV), indicating a broadened CEF ground state. Another experimental evidence for an anomalous $\Gamma_5$ ground state comes from the analysis of the first-order magnetic susceptibility in the paramagnetic range\cite{MorinPrB6}. The triplet ground state is strongly magnetic and, in the case of negligible exchange couplings, would result in a diverging susceptibility when approaching 0 K. However, below 30 K, the magnetic susceptibility is reduced with respect to calculations based on the CEF scheme. Again, this may be indicative of a dynamical splitting of the $\Gamma_5$ level, with features approaching those of slightly separated singlets (Van Vleck susceptibility).
As regards the magnetic entropy in PrB$_6$, reliable data are missing in the literature. Difficulties arise not only from the need for a proper account of the phonon contribution, but also from the two successive antiferromagnetic transitions\cite{McCarthy1980B} at $T_N = 6.9$ K and $T^* = 3.9$ K. Both are first-order processes, which requires an accurate determination of the respective latent heats to define the associated jumps in the magnetic entropy. In the antiferromagnetic range, the reported magnetic structures also point to the role of the displacements of the rare-earth. Below $T_N$, the wave vector is identified as the incommensurate\cite{Burlet1988} $\left\langle \frac{1}{4}-\delta\;\;\frac{1}{4}\;\;\frac{1}{2} \right\rangle$ ($\delta = 0.05$). After the second transition, at $T^* = 3.9$ K, the wave vector locks\cite{Burlet1988} onto the $\left\langle \frac{1}{4}\;\frac{1}{4}\;\frac{1}{2} \right\rangle$ wave vector that, in the RB$_6$ series, characterises the involvement of static displacement waves. This influence of the cage context on the antiferromagnetism of PrB$_6$ is confirmed by X-ray diffraction\cite{Walker2009}, which reveals the associated displacement waves. As in the CeB$_6$ case a thermal expansion anomaly in the paramagnetic range is also reported\cite{Novikov2017}, consisting in a large increase of the expansion coefficient while approaching $T_N$.\newline
To further investigate these aspects of the paramagnetic range of PrB$_6$ and their possible relation with specific cage crystal-field effects, we have performed new specific-heat, magnetic and thermal-expansion measurements on single crystals. The specific-heat analysis focuses on the determination of the latent heats of the magnetic transitions and on an improved accounting for the phonon contribution, in order to define the temperature dependence of the magnetic entropy. The magnetic measurements are intended to verify the first-order susceptibility anomaly previously reported in the low-temperature paramagnetic range. The thermal expansion has been investigated using a high-sensitivity capacitance dilatometer, with the aim of achieving better resolution, with respect to X-ray diffraction, of the volume changes at low temperature. These experimental results are then discussed in relation to the predicted effects of a non-Kramers CEF ground state in the cage context. A following paper will be devoted to a quantitative test of these predictions, based on a model that accounts for a distributed Pr$^{3+}$ ion inside the cage in the paramagnetic range of PrB$_6$.

\section{Experimental determination of the magnetic entropy}
\subsection{Specific heat measurements}
\label{SpecHeat}
\subsubsection{Protocol}
In order to improve the experimental determination of the magnetic entropy of PrB$_6$, new specific heat measurements were performed, with special attention devoted to the latent heat determination at the magnetic transitions, at $T_N$ and  $T^*$. Both the PrB$_6$ and LaB$_6$ samples, the latter being used as a non-magnetic reference, are single crystals grown at the IPMS in Kyiv. The synthesis process is detailed in the supplementary material of reference \cite{Azarevich2024}. Natural boron was used in the synthesis and, due to the mass contrast, one should be aware of possibly significative deviations from highly enriched  $^{11}$B boron samples grown for neutron scattering experiments. The specific heat samples, cut from larger crystals, are platelets with masses $m =$ 6 mg for LaB$_6$ and $m =$ 3.18 mg for PrB$_6$.

\begin{figure}
\includegraphics[width=\columnwidth]{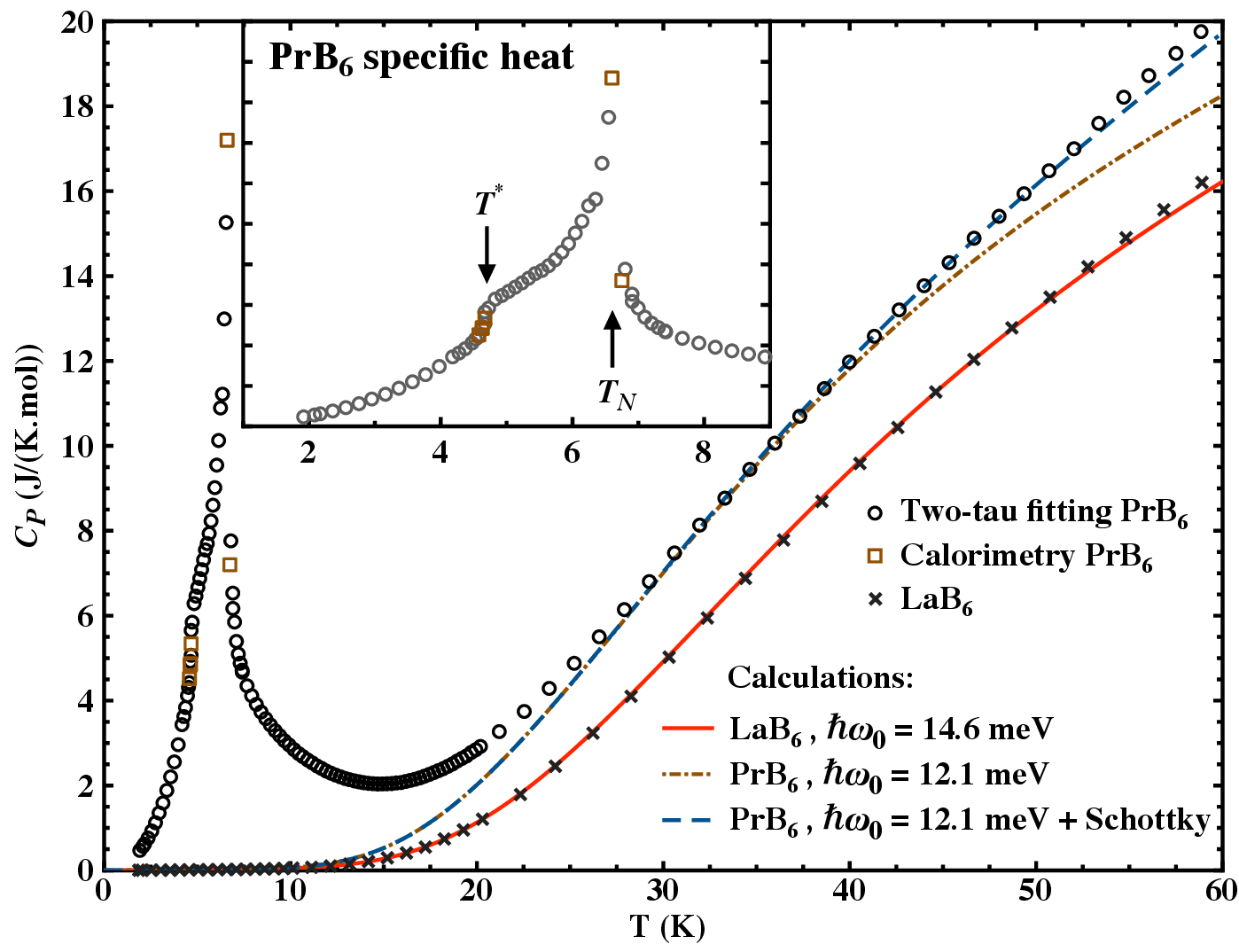}
\caption{\label{SpecHeatData} Experimental specific heat data for PrB$_6$ and baseline curves. The round dots are obtained from the standard two-tau fitting, whereas the square are from the long heating pulses analysis (calorimetry). The inset shows detail of the low temperature range, the arrows pointing to the two magnetic transitions. The non-magnetic reference data for LaB$_6$ are displayed as crosses. The continuous line shows the computed phonons and conduction electrons contributions for LaB$_6$. The line with short dots is the equivalent calculation for PrB$_6$, the longer dots line including the Schottky contribution of excited CEF states. All calculations are for $\gamma_e= 2.3$ mJ/(K$^2\cdot$mol) and  $\hbar  \Omega_{0} = 44.6$ meV (see text).}
\end{figure}

The specific heat measurements were performed using an automated Quantum Design PPMS system, between 1.8 K and 60 K. The thermal coupling between the sample and the setup platform was improved by use of a very small quantity of Apiezon N grease, accounted for in the addenda measurements. Except at the first-order magnetic transitions of PrB$_6$,  at $T_N$ and  $T^*$, the default fitting method of the relaxation process, with two time constants was used. Fig. \ref{SpecHeatData} shows the results of the specific heat measurements up to T = 60 K. The LaB$_6$ non-magnetic reference measurements are the same as published in ref.  \onlinecite{Amara2020}.

One can notice that the LaB$_6$ data cannot be directly used as a non-magnetic background for PrB$_6$. In the paramagnetic range, a simple subtraction of the LaB$_6$ curve from the PrB$_6$ one would leave a substantial offset.  This offset, observed well below the excited $\Gamma _3$ doublet, cannot be attributed to a crystal field Schottky anomaly. As in the CeB$_6$ case, a tailored description of the phonon contribution is necessary (see section \ref{Phonons} below)

\subsubsection{Latent heats determination}
\begin{figure}
\includegraphics[width=\columnwidth]{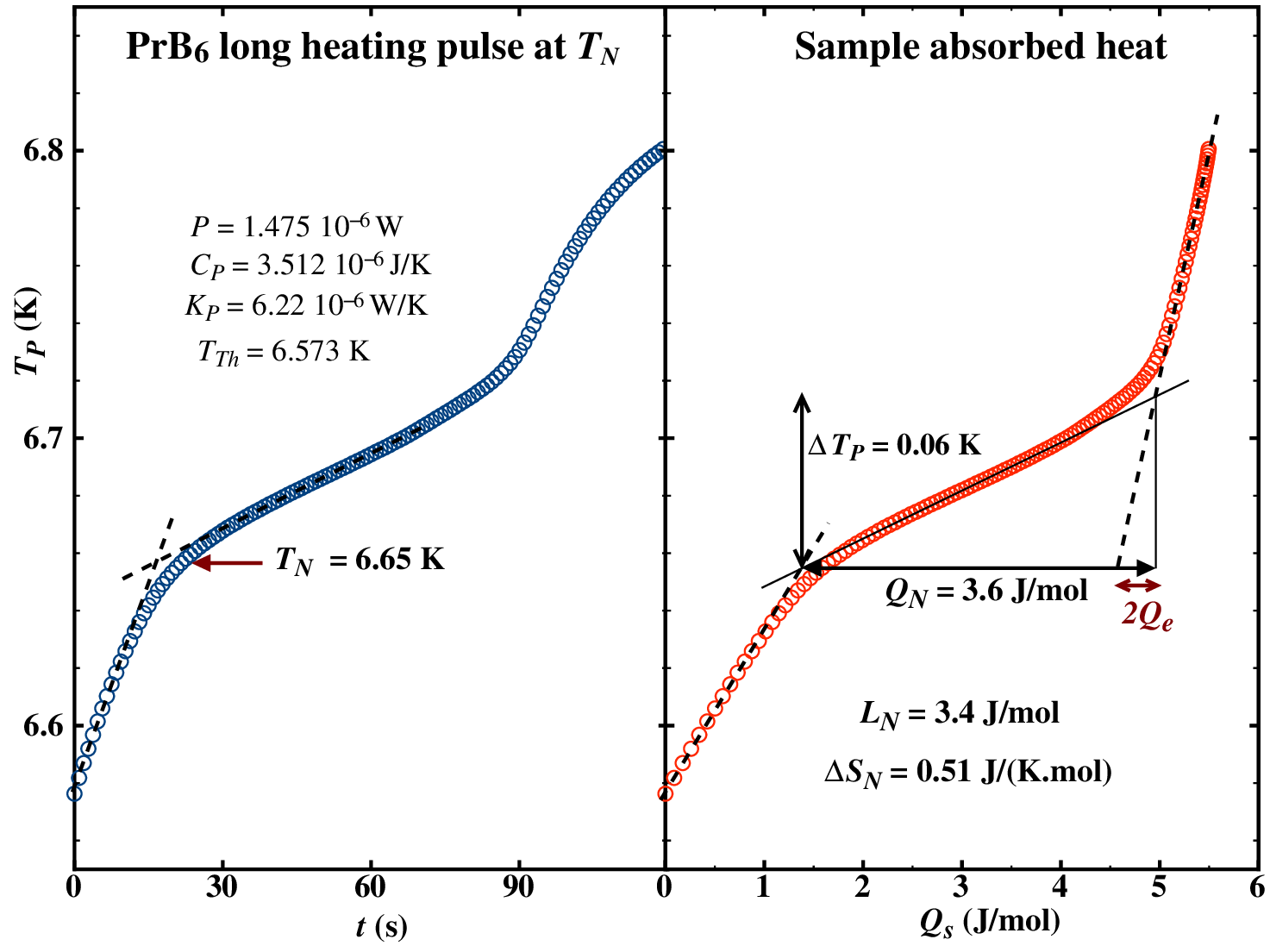}
\caption{\label{PulseTN} Example of a heating pulse at $T_N$. Left part : platform temperature $T_P$ as function of time. Right part: $T_P$ as function of the heat $Q_s$ transferred to the sample, with a graphical determinations of $Q_e$, the total transition heat, and of $Q_e$, the excess heat due to the sample overheating.}
\end{figure}

Once defined the effective transition temperatures $T_N =$ 6.7 K and  $T^* =$ 4.7 K for our PrB$_6$ sample, long heating pulses where used, with adjusted duration and temperature amplitude, in order to observe the first-order transitions "plateaux". The method is detailed if ref. \onlinecite{Amara2020} and allows to define the latent heat required for completing the transition. One assumes that, during the pulse, the sample and the platform, that holds the heater and thermometer (which reads the temperature $T_P$), have only conductive exchange of heat with a stable thermostat at temperature $T_{Th}$. The heat transferred to the sample during a pulse, starting at time $t = 0$, with initially stable temperature $T_P (0) = T_{Th}$, results from a constant heating power $P$ delivered to the platform of specific heat  $C_p$ and a conductive coupling $K_p$ with the thermostat. Within these hypotheses, at time $t$, the heat transferred to the sample reads as:
\begin{equation}
Q_s (t) = \int_{0}^{t}\left[ P-K_p (T_P ( t^{\prime }) - T_{Th}) - C_p \frac{dT_P}{dt}\right]dt^{\prime}
\label{HeatEq}
\end{equation}
Fig. \ref{PulseTN} shows the example of one heating pulse that covers the transition at $T_N$. The heat $Q_s$ transferred to the sample, that results from a numerical integration of Eq. \ref{HeatEq} appears on the right. All of the required parameters are displayed. During the transition, the sample is not homogenous in temperature, hence the slope in the supposed plateau. A the end of the transition, the sample is, on average, overheated (see Ref. \onlinecite{Amara2020}), the temperature at the platform reaching $T_N + \Delta T_P$. The total heat transferred to the sample to complete the transition is $Q_N$ (with a $Q_N =$ 3.6 J/mol value from the graph), which includes the overheating $Q_e$. Within the simplest assumptions, the average sample overheating is $\Delta T_P /2$ and the excess heat $Q_e$ reads as:
\begin{equation*}
Q_e= C_s (T_{N_+}) \frac{\Delta T_P} {2}
\end{equation*}
where $C_s (T_{N_+})$ is the sample specific heat immediately above the transition temperature.  As the slope of $Q_s (T_P)$ after the plateau represents the inverse of the specific heat $C_s (T_{N_+})$, $Q_e$ can be estimated from the graph. Here, one reads $Q_e \approx 0.2$ J/mol.

\begin{figure}
\includegraphics[width=\columnwidth]{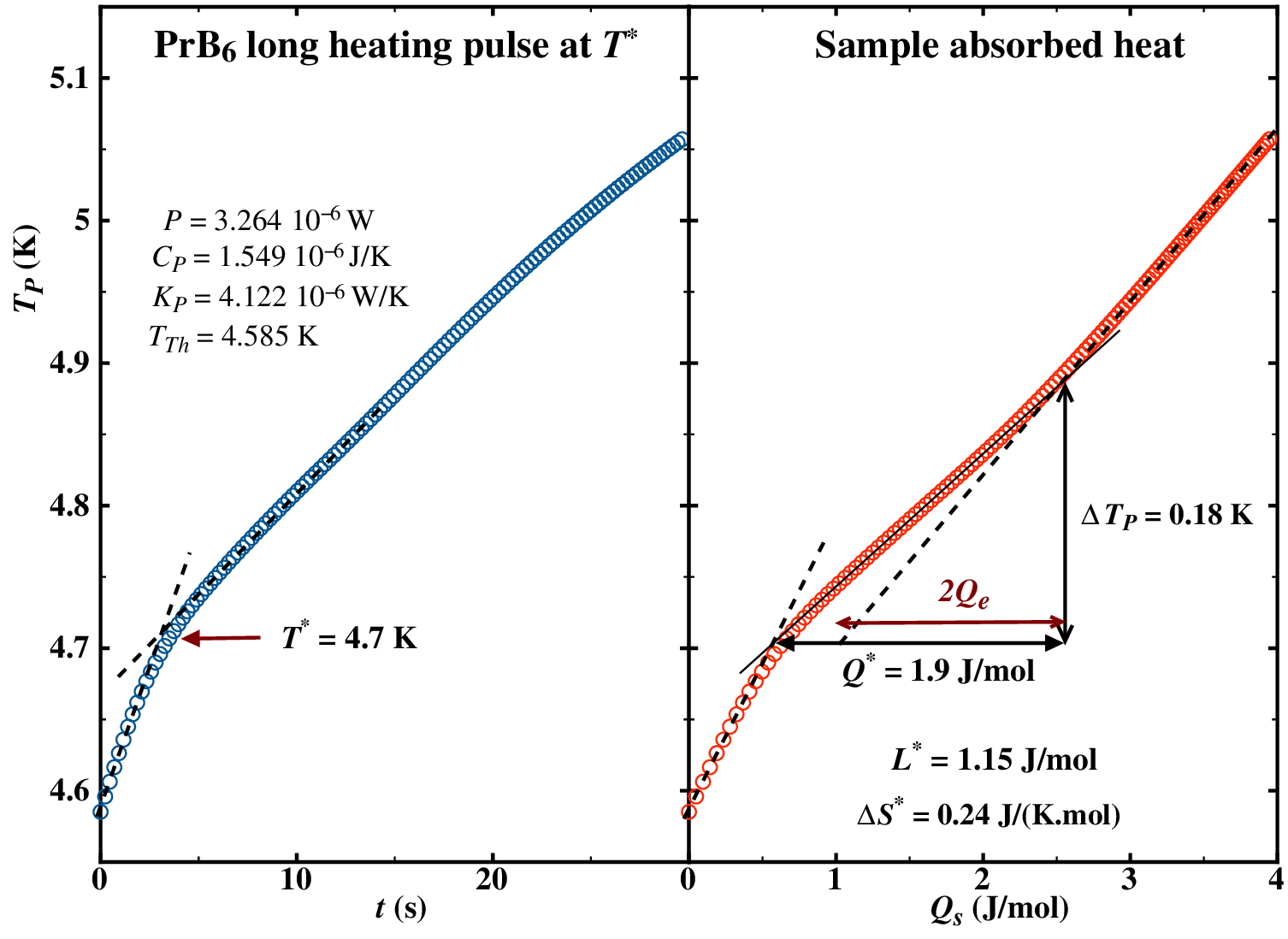}
\caption{\label{PulseT*} Example of a heating pulse at $T^*$. Left part: platform temperature $T_P$ as function of time. Right part: $T_P$ as function of the heat $Q_s$ transferred to the sample. $Q^*$ is the total heat transferred during the transition process, whereas $Q_e$ is the excess heat due to the sample overheating. }
\end{figure}

From this analysis and that of additional pulses, the retained value for the latent heat at $T_N$ and its estimated error are:
\begin{equation*}
L _N = 3.3 \pm 0.3\;  \text{J/mol}
\end{equation*} 
The corresponding step in entropy at $T_N=$ 6.65 K is:
\begin{equation*}
\Delta S_N = L_N /T_N = 0.5 \pm 0.05\;  \text{J/(K mol)}
\end{equation*}
The analysis of the heating pulses at $T^*$ are more delicate. As shown in the example of Fig. \ref{PulseT*}, due to the lower latent heat involved and larger specific heat in the antiferromagnetic state, the "plateau" is even less pronounced than at $T_N$, with an overheating $\Delta T_P = 0.18$ K three times larger. The graphic estimate for the excess heat $Q_e$ transferred to the sample during the transition process is 0.75 J/mol. From several such measurements and analyses, one obtains the latent heat:
\begin{equation*}
L ^* = 1.2 \pm 0.4\;  \text{J/mol}
\end{equation*} 
The change in entropy at $T^*=$ 4.7 K is then:
\begin{equation*}
\Delta S^* = L ^* /T ^* =  0.27\pm 0.07\;  \text{J/(K mol)}
\end{equation*}

%\begin{figure}
%\includegraphics[width=13.5cm]{Fig2.eps}
%\caption{\label{LongPulses} Upper part: Thermal exchange scheme during a first-order transition in heating conditions. The sample's platform has heat capacity $C_P$, thermal resistance to the thermostat $K_P$ and absorbs a heating power $P$. As the transition front moves, an increase in the platform temperature $T_P$ develops. Lower part: Two examples of $T_P (t)$ temperature profiles, showing the crossing of the antiferromagnetic transition in CeB$_6$, when using long heating pulses. These processes have the same duration but differ in their starting temperature and temperature rise (related to the heating power $P$). The full lines are adjusted considering exponential relaxation laws above and below a linear "plateau" of duration $\Delta t$ and temperature rise $\Delta T_P$. }
%\end{figure}

\subsection{Phonons contribution to the specific heat}  \label{Phonons}
After determining the latents heats and associated entropy steps, defining the temperature variation of CeB$_6$ magnetic entropy requires proper identification of the non-magnetic contributions. This step, as is standard, relies on the specific heat data for a non-magnetic element in the series, LaB$_6$ in the present case (lower curve in Fig. \ref{SpecHeatData}).
In a non-magnetic metal, the specific heat is well described considering two contributions, $C_{e}$ from the conduction electrons, $C_{ph}$ from the phonons:
\begin{equation}
C(T) = C_{e}(T) +C_{ph}(T)
\end{equation}
In Refs. \onlinecite{Amara2019, Amara2020}, we introduced our description of the phonons contribution to the low temperature specific heat. The RB$_6$ lattice is schematized as a harmonic cubic lattice consisting of two elements, the cages, of mass $M$, that realize an elastic matrix, and the cage guest, of mass $m$. Inside the cage, the guest movement is that of a harmonic oscillator with frequency $\omega_0$. This model, yields two phononic branches for a given polarization. These are representative of the two lowest branches of the rare-earth hexaborides. As these branches share quantitative (the $q = 0$ gap) and qualitative (the overall shape, with, in particular, the flattened lower branch) traits, a  further simplification consists in ignoring the polarization dependence of the phonons.  For a given phonon with wave vector $\mathbf{q} =  \left[ q_x\;q_y \;q_z \right]$ (the reciprocal cubic lattice $q_x,\;q_y, \;q_z$ components are here dimensionless), instead of three polarizations dependent frequencies, only one,  an averaged representative, is considered in this model. Along the reciprocal space direction defined by $\mathbf{q} $, the first Brillouin zone limit is reached for a vector $\mathbf{\Gamma B_q}=\gamma \mathbf{q}$, where $\mathbf{\Gamma}$ is the cubic zone center, $\mathbf{B_q}$ a point on the zone border and $\gamma$ a positive scalar. Along this direction, the minimal spacing of perpendicular, direct space lattice planes is $ d = a/2 \| \mathbf{\Gamma B_q}\|$, were $a$ is the cubic lattice parameter. The shift in the phase of the phonon wave, between adjacent lattice planes, is then:

\begin{equation*}
\Delta \varphi_\mathbf{q} = \frac{2 \pi}{a} \| \mathbf{q}\| d = \pi\frac{ \| \mathbf{q}\|}{\| \mathbf{\Gamma B_q}\| }= \frac{\pi}{\gamma}
\end{equation*}

Thanks to the cubic shape of the first Brillouin zone,  $\Delta \varphi_\mathbf{q}$ can be simply expressed using $\mathbf{q}$ indices. For instance, for a vector $\mathbf{q}=\left[ q_x , q_y , q_z \right]$ with all indices positive, $q_x$ being the largest, the intercept of the direction of $\mathbf{q}$ with the zone border occurs at $\mathbf{\Gamma B_q} =\gamma \mathbf{q} = \left[ \frac{1}{2} , \gamma\; q_y , \gamma\; q_z \right]$. This allows to identify $\gamma = \frac{1}{2 q_x}$ and it follows that:
\begin{equation*}
\Delta \varphi_\mathbf{q} = 2 \pi q_x
\end{equation*}
Then, writing down the classical equations of motion, one obtains the two frequencies, $\omega_{+}(\mathbf{q})$ or $\omega_{-}(\mathbf{q})$, that obey the equation:
\begin{equation}
\label{EqDisp1}
\cos ( 2 \pi q_x) = 1 - 2 (1 + \frac{m}{M} \frac{{{\omega _0}^2{\kern 1pt} }}{{{\omega _0}^2 - {\kern 1pt} {{\omega_{\pm}(\mathbf{q})}^2}}})\frac{{{\omega_{\pm}(\mathbf{q})^2}}}{{{\Omega_0}^2}}
\end{equation}
where $\Omega_0$ is an average, over the three polarizations and all reciprocal space directions, of the acoustic frequency at the border zone for a lattice of supposedly empty cages. 
This allows to define the two lowest phonon branches using only two frequency parameters, $\omega _0$ and $\Omega_0$. Introducing the mass ratio $\alpha = m/M$, the frequency ratio $\rho = \Omega_{0}/ \omega_{0}$ and the function $K(\mathbf{q}) = {\rho^2} (1- \cos(2\pi q_\mu))/2$ (where $\mu$ refers to the $\mathbf{q}$ index with largest absolute value):
\begin{equation}
\label{EqDisp2}
{\omega_{\pm}(\mathbf{q})} = \omega_{0} \sqrt {\frac{{1 + \alpha  + K(\mathbf{q}) \pm \sqrt {{{(1 + \alpha + K(\mathbf{q}))}^2} - 4 K(\mathbf{\mathbf{q}})} }}{2}}
\end{equation}
where the $+$ and $-$ options respectively give the expression for the optic and acoustic branches.

Once defined the two lowest branches phonons frequencies, for any wave vector and polarization, the phonons contribution to the specific heat at low and intermediate temperatures reads as:
\begin{equation}
 C_{ph} (T) = 3\sum\limits_{\mathbf{q}, \pm} {k_{\text{B}}{{\left( {\frac{{\hbar \,\omega_{\pm}(\mathbf{q})}}{{k_{\text{B}}T}}} \right)}^2}\frac{{{e^{ - \frac{{\hbar \,\omega_{\pm}(\mathbf{q})}}{{k_{\text{B}}T}}}}}}{{{{\left( {1 - {e^{ - \frac{{\hbar \,\omega_{\pm}(\mathbf{q})}}{{k_{\text{B}}T}}}}} \right)}^2}}}} 
\end{equation}
where the sum covers all nodes $\mathbf{q}$ within the first Brillouin zone.  Obviously, in practical calculations, this becomes a discrete sum over numerous samples inside a representative volume\cite{Amara2020} of the first Brillouin zone.

\subsection{Magnetic entropy extraction}
The description of the non-magnetic contribution to the specific heat of PrB$_6$ involves three parameters, the  previously defined frequencies $\omega_{0}$ and $\Omega_{0}$, and the coefficient $\gamma_e$ for the electronic linear contribution at low temperature, given by $C_{e}(T) =\gamma_e T$ . To estimate these parameters, we first refer to values for the non-magnetic reference LaB$_6$.

The coefficient for the electronic contribution in LaB$_6$,  $\gamma_e = 2.3\pm0.1$ mJ/(K$^2$.mol), can be directly obtained from a plot of $C_P /T$ versus $T^2$. 
From inelastic neutron scattering (INS) on LaB$_6$, there is available quantitative data for $\omega_{0}$. Using our phonons model, the most direct determination of $\omega_{0}$ is from the zone center, where $K(\mathbf{q})$ cancels, and the optical branch energy is $E_O^{\Gamma} = \hbar \omega_0 \sqrt{1+\alpha}$.  From Ref. \onlinecite{Smith1985} one has $E_O^{\Gamma}=24$ meV at room temperature, which yields $\hbar \omega_0 = 13.6$ meV, using the mass ratio $\alpha$ for 98\% $^{11}$B-enriched boron.

From this initial $\omega_0$ value, one can proceed with a least squares fit to the experimental LaB$_6$ specific heat data, by adjusting $\omega_0$ and $\Omega_0$.  Using the full temperature range, the refinement yields $\omega_0$= 15.7 meV and $\Omega_0$= 38.6 meV. This $\omega_0$ value is significantly larger than the INS determination and, whereas the model is inherently a low-temperature model, the description below 20 K is unsatisfactory. This shows that the model shortcomings, at higher temperatures, are compensated for by an increase in $\omega_0$. If one restricts the fit to data below 20 K, the low-temperature description improves substantially, with limited discrepancies at higher temperatures. The resulting fit, represented by the full line in Fig. \ref{SpecHeatData}, provides refined values of $\hbar \omega_0=14.6$ meV (still 7 \% larger than the INS determination) and $\hbar \Omega_0=44.6$ meV.

Starting from the LaB$_6$ values, one can then proceed, adjusting parameters for PrB$_6$. Given that the low-temperature specific heat data for PrB$_6$ is dominated by magnetic contributions, a direct determination of the conduction electrons coefficient $\gamma_e$ is not possible. As an electron mass enhancement is expected in the paramagnetic range of a magnetic element, the $\gamma_e$ value for PrB$_6$ should exceed that of LaB$_6$. However, in the absence of an independent determination, the lower-bound LaB$_6$ value, $\gamma_e= 2.3$ mJ/(K mol), will be used for computing the non-magnetic specific heat of PrB$_6$.

As it can be interpreted as the average upper acoustic frequency for a lattice of empty boron cages (i.e., a lattice where the mass of the lanthanide is nullified while retaining all other properties), $\Omega_{0}$ is the frequency parameter that should vary the less across the series.  For instance, for a longitudinal wave along a fourfold direction, the corresponding $\Omega_{11}^{La}$ top frequency for LaB$_6$ would read as: $\Omega_{11}^{La} = \sqrt{\frac{a_{La} \cdot C_{11}^{La}}{M}}$ where $a_{La}$ is LaB$_6$ cubic lattice parameter,  $C_{11}^{La}$ the related cubic elastic constant and $M$ the mass of the B$_6$ cage. From the analogous expression for PrB$_6$, one obtains the ratio: $\frac{\Omega_{11}^{Pr}}{\Omega_{11}^{La}}=\sqrt{\frac{ a_{Pr} \cdot C_{11}^{Pr}}{a_{La} \cdot C_{11}^{La}}}$. As lattice parameters and elastic constants values are available in the literature, one can then extrapolate $\Omega_{11}^{La}$ from $\Omega_{11}^{Pr}$. Using the low temperature $C_{11}$ values from ref. \onlinecite{Nakamura1994} and lattice parameters from ref. \onlinecite{Samsonov1963} yields $\Omega_{11}^{Pr} \approx 0.993\; \Omega_{11}^{La}$. This supports the assumption that $\Omega_{0}$ changes minimally between LaB$_6$ and PrB$_6$. 

Consequently, the value\footnote{Note that the $\Omega_{0}$ value published as the result of a previous fit of the LaB$_6$ data in Ref.~\onlinecite{Amara2020} is wrong due to an error in the calculation code. Except for the fit yielding this erroneous value of $\Omega_{0}$, this error has no relevant consequence on the description of the specific-heat baseline} $\hbar \Omega_{0} = 44.6$ meV, derived from the LaB$_6$ fit, is retained for PrB$_6$.
As regards the neutron phonon spectroscopy on PrB$_6$, no data are available from which to extract the energy $E_O^{\Gamma}$ of the optical phonon. From the dispersions of the acoustic branch\cite{Alekseev2015, Kohgi2006}, two values are obtained for the energy of longitudinal acoustic phonons at the $X$ point of the zone border: 10.2 meV\cite{Alekseev2015} and 10.7 meV\cite{Kohgi2006}. Using our simplified dispersions, these energies are identified with $E_A^X$, which depends on both $\omega_{0}$ and $\Omega_{0}$:
\begin{equation}
\label{EAX}
E_A^X = \hbar {\omega _0}\sqrt {\frac{{1 + \frac{m}{M} + \frac{{{\Omega _0}^2}}{{{\omega _0}^2}} - \sqrt {{{(1 + \frac{m}{M} + \frac{{{\Omega _0}^2}}{{{\omega _0}^2}})}^2} - 4\frac{{{\Omega _0}^2}}{{{\omega _0}^2}}} }}{2}}
\end{equation}
Despite the assumption that $\Omega_{0}$ shows only slight variations across the RB$_6$ series, the value obtained from LaB$_6$ need a renormalisation to account for the boron mass effect: the samples used in neutron scattering are highly enriched in $^{11}$B, whereas our specific-heat specimens contain natural boron. The molar mass for a 98\% $^{11}$B-enriched B$_6$ cage is $M^* = 65.93$ g, versus $M = 64.86$ g for natural boron. The corrected $\Omega_{0}$ value for the $^{11}$B neutron sample is given by $\Omega_{0}^* = \sqrt{\frac{M}{M^*}} \Omega_{0}$, which yields a moderately adjusted value of $\hbar \Omega_{0}^* = 44.2$ meV for a PrB$_6$ sample 98\% enriched in $^{11}$B. Eventually, using Eq.~(\ref{EAX}), the $E_A^X$ values in Refs.~\onlinecite{Alekseev2015, Kohgi2006} yield estimates for $\hbar\omega_{0}$ in PrB$_6$ of 10.9 and 11.5 meV, respectively.

\begin{figure}
\includegraphics[width=\columnwidth]{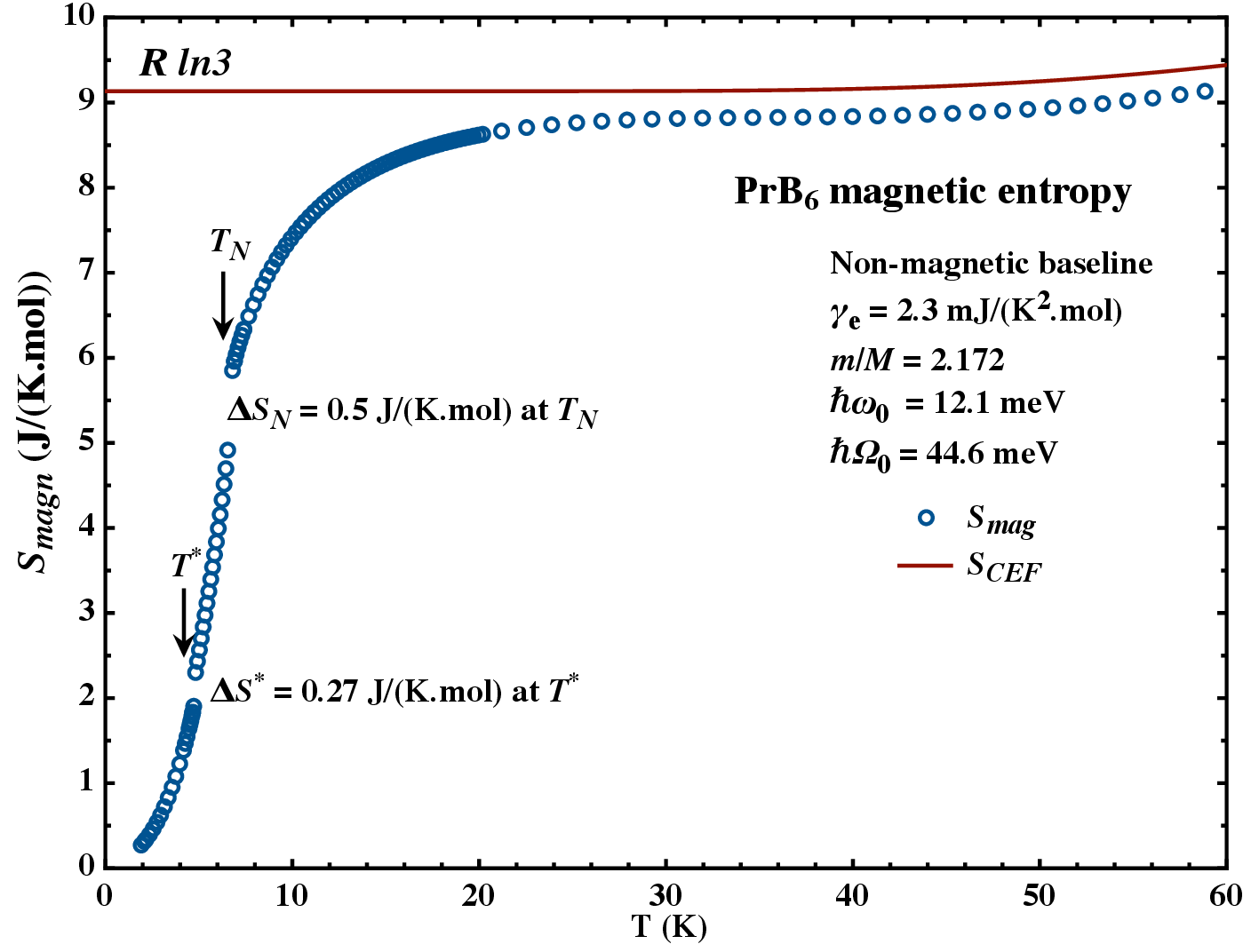}
\caption{\label{Entropy} Temperature dependence of PrB$_6$ $S_{mag}$ magnetic entropy as obtained from the specific heat data, after subtraction of the non-magnetic baseline, and including the latent heat steps at T$^*$ and T$_N$. The continuous line is the expected CEF paramagnetic entropy $S_{CEF}$, computed using the level scheme in Ref. \onlinecite{Loewenhaupt1986}.}
\end{figure}

Using these parameters to compute the non-magnetic baseline for PrB$_6$ produces curves that overshoot the experimental data above 25 K. Adjusting the most uncertain parameter $\omega_{0}$ to $\hbar\omega_{0} = 12.1$ meV suppresses this overshoot (see the line with shorter dashes in Fig.~\ref{SpecHeatData}). However, above 40 K, the calculation falls significantly below the experimental data. This can be ascribed to the contribution of the Schottky anomaly caused by the excited CEF levels, the $\Gamma_3$ doublet at 314 K and the $\Gamma_4$ triplet at 377 K (see Ref.~\onlinecite{Loewenhaupt1986}). Indeed, including the contributions of the excited CEF states to the specific heat (Fig.~\ref{SpecHeatData}, longer-dashed line) significantly improves the description above 40 K. The remaining discrepancies, around 60 K, are in line with those observed in LaB$_6$ and reflect the limitations of the phonon model, which neglects the influence of higher-energy phonon bands. This baseline, which best describes the PrB$_6$ data when including the Schottky anomaly, corresponds to the following set of parameters:

$m/M = 2.172$, $\gamma_e = 2.3$ mJ/(K$^2\cdot$mol), $\hbar \Omega_{0} = 44.6$ meV and $\hbar\omega_{0} = 12.1$ meV.

The optimal $\omega_{0}$ value is larger than those derived from inelastic neutron scattering experiments\cite{Alekseev2015, Kohgi2006}, in similar proportions as for LaB$_6$. While neutron spectroscopy is a direct determination of $\omega_{0}$, the value used here is a fitting parameter that defines a rather simple phonon model: differences are to be expected.

Nevertheless, the optimized $\omega_{0}$ value effectively serves its purpose, which is to reproduce the phonon baseline and, by subtraction, to isolate the magnetic contribution $C_{mag}$ to the specific heat. For temperatures below 1.9 K, the missing specific-heat experimental data are replaced by an extrapolation using a fit over the low-temperature magnetic phase (the function of the temperature used includes a linear term and an adjustable power-law term). One can then proceed with the determination of the magnetic entropy $S_{mag}(T)$, using numerical integration:
\begin{equation}
 S_{mag} (T) = \int_{0}^{T^*} \frac{C_{mag}(T^{\prime})}{T^{\prime}} dT^{\prime}+\Delta S^*+ \int_{T^*}^{T_N} \frac{C_{mag}(T^{\prime})}{T^{\prime}} dT^{\prime}+\Delta S_N+\int_{T_N}^{T} \frac{C_{mag}(T^{\prime})}{T^{\prime}} dT^{\prime}
\end{equation}
where $\Delta S^*$ and $\Delta S_N$ are the entropy steps deduced from the latent heats of the magnetic transitions. The results for the chosen non-magnetic baseline are shown in Fig.~\ref{Entropy}, along with the temperature variation of $S_{CEF}$, the CEF entropy associated with the level scheme of Ref.~\onlinecite{Loewenhaupt1986}. Above 30 K, $S_{mag}$ closely follows $S_{CEF}$, with a small offset of approximately 0.3 J/(K$\cdot$mol). This discrepancy is not surprising, since $S_{mag}$ includes cumulative errors from the extrapolation below 2 K and the determinations of the latent heats at $T^*$ and $T_N$. In particular, the determination of $\Delta S^*$ is quite delicate, with its uncertainty likely underestimated: some of the latent heat may have been lost in the experimental data analysis.
Despite the offset in the paramagnetic value of $S_{mag}$, the following conclusions can be drawn:
\begin{itemize}
\item[-] above 30 K, the experimental magnetic entropy is consistent with the published CEF level scheme, approaching the characteristic $R \ln 3$ J/(K$\cdot$mol) value of the triply degenerate $\Gamma_5$ ground state, with visible Schottky contributions from the excited CEF levels above 40 K.
\item[-] in the paramagnetic range, the magnetic entropy increases rapidly below 30 K, from about the doublet $R \ln 2$ J/(K$\cdot$mol) value immediately above $T_N$ to the evolving $R \ln 3$ J/(K$\cdot$mol) CEF triplet value above 30 K.
\end{itemize}
The substantial loss of magnetic entropy in the paramagnetic range, below 30 K, while approaching $T_N$, is reminiscent of the CeB$_6$ case.\cite{Amara2020}

\section{Magnetic measurements}
\label{Magn}
\begin{figure}
\includegraphics[width=\columnwidth]{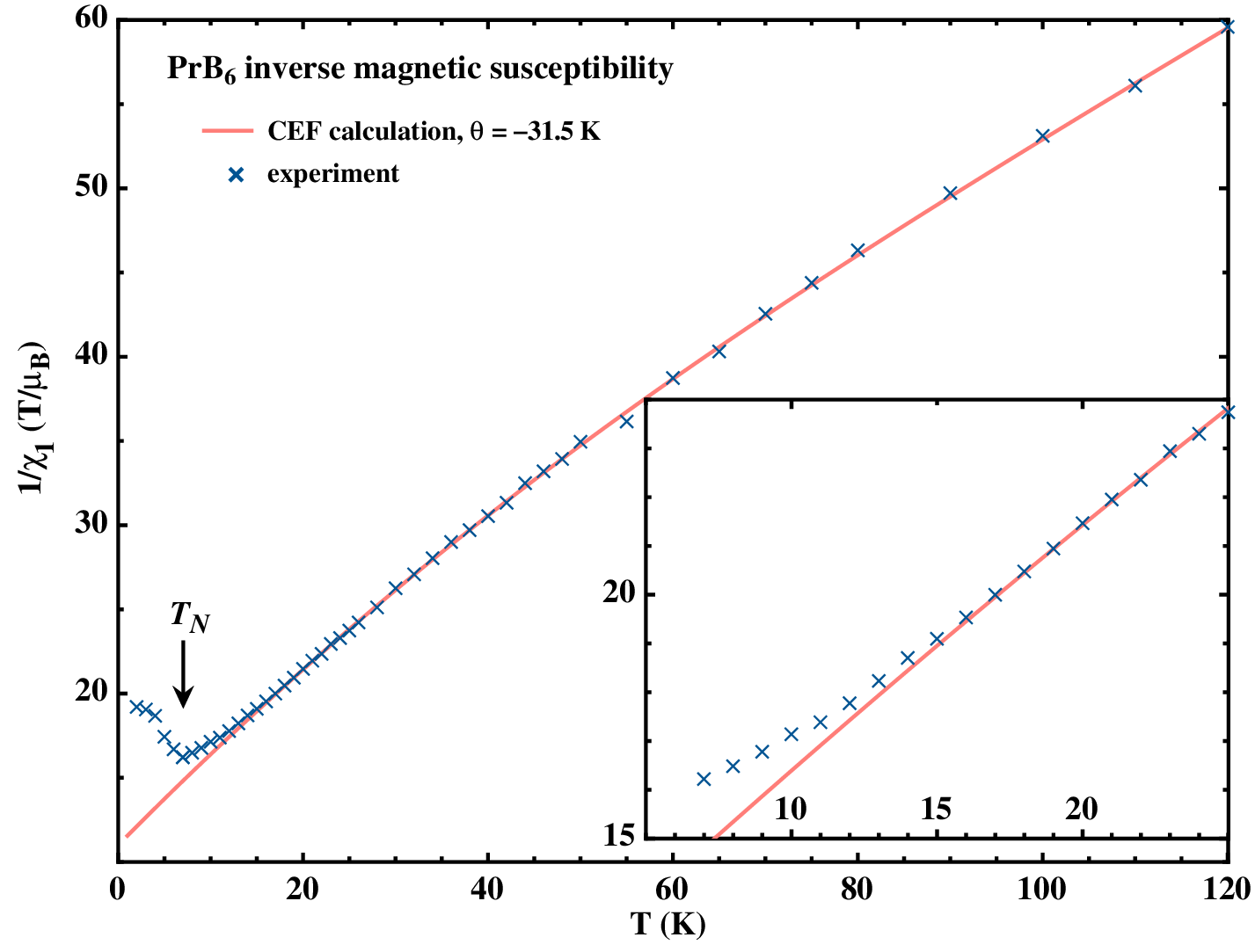}
\caption{\label{InvChi} Temperature dependence of PrB$_6$ inverse first-order magnetic susceptibility per atom, obtained from Arrott plots. The line is the result of CEF calculations using the scheme from Ref. \onlinecite{Loewenhaupt1986} with exchange interactions ($\theta$ = -31.5 K) visually adjusted to fit the experimental data. The inset shows detail of the low temperature paramagnetic range.}
\end{figure}
The first-order magnetic susceptibility measurements reported in Ref.~\onlinecite{MorinPrB6} exhibit a peculiar low-temperature behaviour that contradicts the CEF scheme of Ref.~\onlinecite{Loewenhaupt1986}. Below 30 K, the data show an inflection recalling that of a non-magnetic CEF ground state, whereas neutron spectroscopy identifies a strongly magnetic $\Gamma_5$ triplet ground state.
To investigate whether this anomaly is intrinsic to PrB$_6$, magnetic measurements were undertaken on a sample from a different source: a single crystal from the same batch as the one used for our specific-heat measurements. The sample was not oriented, since the cubic symmetry imposes an isotropic first-order susceptibility. Magnetization data were collected using the institute's extraction magnetometer, at constant temperature and variable field up to 9 T, in the temperature range from 2 to 200 K. Corrections were applied for the diamagnetic contribution of the acrylic sample holder, and the first-order susceptibility was extracted using an automated method inspired by the Arrott plot technique\cite{Arrott1957}. The results are presented in Fig.~\ref{InvChi}, with a detailed view of the low-temperature region shown in the inset.
The continuous line shows CEF calculations according to the scheme of Ref.~\onlinecite{Loewenhaupt1986}, the only visually adjusted parameter, which shifts the calculated curve vertically, being the exchange interaction temperature $\theta = -31.5$ K. These results are consistent with those of Ref.~\onlinecite{MorinPrB6}, showing a negligible difference of 0.5 K in the adjusted $\theta$ value. Compared with CEF calculations according to the scheme of Ref.~\onlinecite{Loewenhaupt1986}, the data confirm the anomalous behaviour in the low-temperature paramagnetic range: below 20 K, the measurement deviates from the CEF curve, in contradiction with the magnetic nature of the $\Gamma_5$ CEF ground state. Therefore, some mechanism is at play that reduces the first-order magnetic susceptibility at low temperature.

\section{Thermal expansion}
\label{Thexp}
\subsubsection{Experimental setup}
Thermal expansion measurements were performed using the institute's magnetostriction setup, which features a non-magnetic copper-beryllium capacitance dilatometer, sensitive to changes in the sample length along a horizontal direction. This dilatometer is housed within a helium flow cryostat, enabling measurements from 2 to 300 K. The cryostat is equipped with a superconducting magnet capable of generating a horizontal magnetic field of up to 6 T. The dilatometer can be rotated about a vertical axis, allowing length measurements for any magnetic field direction within the horizontal plane.
Using Stycast 2850FT, the sample is glued onto a custom-sized Cu-Be sample holder, and this assembly is then inserted into the capacitance cell. The free end of the sample is pressed against the moving electrode, converting length changes into capacitance variations. The cell and associated capacitance bridge are capable of detecting length changes as small as 1 \AA: for samples a few millimetres in length, relative expansions down to $10^{-7}$ can be resolved. For the present study, the system was used to measure thermal expansion in a constant, zero magnetic field. As the temperature varies, the measured capacitance includes large contributions from the copper-beryllium cell, which, moreover, exhibits some thermal hysteresis. To account for these effects, a reference sample, made of the same copper-beryllium alloy as the cell, is measured following exactly the same temperature cycle and rate as the investigated sample.

The thermal expansion measurement yields the temperature dependence of the sample length $l(T)$. The linear thermal expansion coefficient $\alpha (T)$ is then obtained through a numerical approximation of the derivative:
\begin{equation}
\alpha(T) = \frac{1}{l} \left. \frac{\partial l}{\partial T} \right)_{P}
\end{equation}
To enhance the temperature resolution, one usually resorts to small intervals (less than 1 K). This approach can result in a noisy numerical derivative, as the small temperature increments produce length changes that approach the detection limit of the system. Additionally, the sample's contact point on the moving electrode may occasionally slip, introducing discontinuities in $l(T)$ and associated peaks in $\alpha (T)$. After the numerical derivative, some level of low-pass filtering is required to mitigate the dispersion in $\alpha (T)$ (see Ref.~\onlinecite{Amara2025}).

\begin{figure}
\includegraphics[width=\columnwidth]{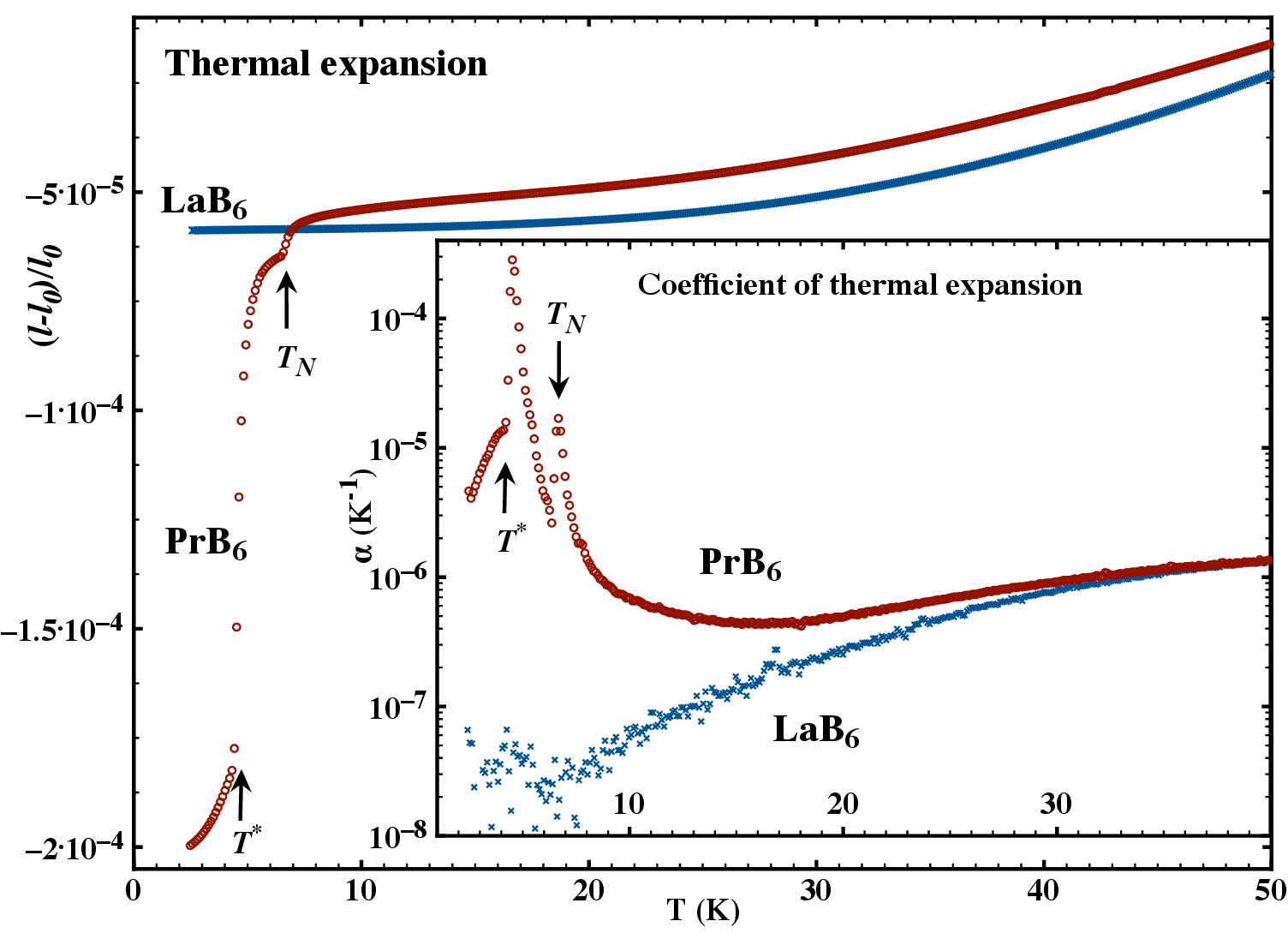}
\caption{\label{ThermExp} Thermal expansion of LaB$_6$ and PrB$_6$ as the relative changes in length with respect to a 60 K reference length $l_0$. The samples lengths are measured while cooling down from 60 K to 2.5 K at a 0.2 K/min rate. The inset shows the deduced linear thermal expansion coefficients $\alpha$, as unfiltered numerical derivatives. Due to the large contrast of amplitude between the $\alpha$ variations for LaB$_6$ and PrB$_6$, the inset vertical axis is logarithmic.}
\end{figure}

\subsubsection{Measurements}

Here again, a LaB$_6$ single crystal was used as a non-magnetic reference. These LaB$_6$ measurements are those already reported in Ref.~\onlinecite{Amara2025}. As regards the PrB$_6$ data, they were collected using the same single crystal as the one used for the magnetization measurements. The sample was not oriented, but was glued so that its larger dimension, of 5.32 mm, corresponds to the probed length. Figure~\ref{ThermExp} shows the measured relative changes in length for both LaB$_6$ and PrB$_6$, as obtained under zero magnetic field and decreasing temperature between 60 K and 2.5 K at a rate of 0.2 K/min. In the inset, the corresponding linear thermal expansion coefficients, obtained from simple direct numerical differentiation, are displayed using a logarithmic vertical scale. The temperature variations for PrB$_6$ show dramatic discontinuities at $T_N$ and $T^*$, reflecting the two first-order magnetic transitions. In the paramagnetic range, below 30 K (see the inset of Fig.~\ref{ThermExp}), the thermal expansion of PrB$_6$ is well separated from that of the non-magnetic LaB$_6$ reference, with a three-orders-of-magnitude difference close to $T_N$.

\begin{figure}
\includegraphics[width=\columnwidth]{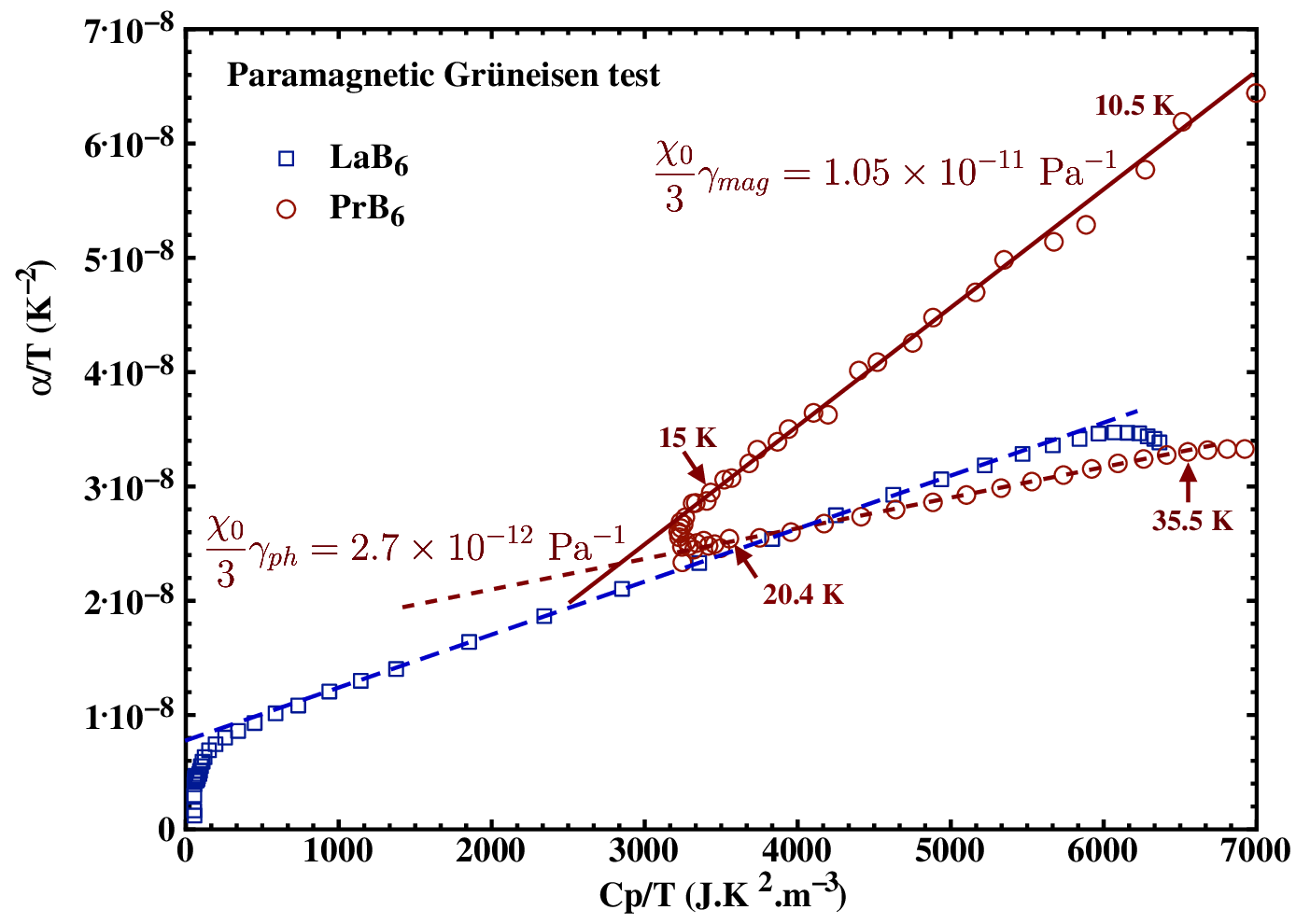}
\caption{\label{GrunAlphaCp} Paramagnetic test of the Gr\"uneisen rule for PrB$_6$ and LaB$_6$, plotting $\alpha /T$ as a function of $C_p /T$ as in Ref. \onlinecite{Amara2025}. This kind of plot highlights the non-electronic, dominant contributions to thermal expansion and specific heat, whose slopes defines a local Gr\"uneisen parameter. Here, in the PrB$_6$ case, $\gamma_{ph}$ is associated with the flat phonons acoustic branch, whereas $\gamma_{mag}$ points to a magnetic effect absent in LaB$_6$. $\chi_0$ is the low temperature lattice compressibility.}
\end{figure}

If only phonon and conduction-electron contributions were involved in the paramagnetic thermal expansion, the temperature dependences of LaB$_6$ and PrB$_6$ would be similar down to the PrB$_6$ ordering temperature. It clearly appears that sizeable differences develop below 30 K. In Ref.~\onlinecite{Amara2025}, we defined an effective Gr\"uneisen parameter, actually the function of the temperature $\gamma_{eff}(T) = (3/\chi_0)\cdot \alpha(T)/C_p(T)$, where $\chi_0$ is the low-temperature compressibility of the lattice. Changes, with temperature, in the dominant contribution to the specific heat and thermal expansion are reflected in the evolution of $\gamma_{eff}$. This is spectacularly illustrated in Fig.~\ref{GrunAlphaCp}, where a plot testing the Gr\"uneisen rule as $\alpha /T$ as a function of $C_p /T$ is displayed, comparing PrB$_6$ with LaB$_6$. In the case of PrB$_6$, at temperatures higher than 20 K (the crossover is at about $C_p /T = 3200$ J\,K$^{-2}$\,m$^{-3}$ and $\alpha /T = 2.8\times10^{-8}$ K$^{-2}$, which correspond to a temperature of 18 K), one observes that the circular PrB$_6$ dots align, similarly to the line of square dots in LaB$_6$, which was ascribed to the flattened acoustic phonon branch\cite{Amara2025}. However, between 15 K and 20 K, a drastic change occurs, with the emergence of a paramagnetic regime clearly distinct from that of the phonons. This means that, below 20 K, there is a dominant magnetic contribution to the thermal expansion, hence the $\gamma_{mag}$ naming for the associated Gr\"uneisen parameter in the figure.
From the data in Ref.~\onlinecite{Nakamura1994}, the low-temperature estimate of the PrB$_6$ lattice compressibility is $\chi_0 = 5.4\times 10^{-12}$ Pa$^{-1}$, a value very close to that for LaB$_6$\cite{Amara2025}. From this, one can derive values for the local Gr\"uneisen parameter corresponding to the slopes in Fig.~\ref{GrunAlphaCp}: $\gamma_{ph} = 1.5$ for the flattened acoustic phonon branch, and $\gamma_{mag} = 5.8$ for the low-temperature paramagnetic range. The concept of the Gr\"uneisen parameter emerges from the anharmonic treatment of the phonon contribution and, a priori, should not apply to a magnetic contribution. However, the large magnetic value $\gamma_{mag} = 5.8$ attests to the tight relation between the volume and the magnetic entropy in the low-temperature paramagnetic range.

\section{Discussion}

\begin{figure}
\includegraphics[width=\columnwidth]{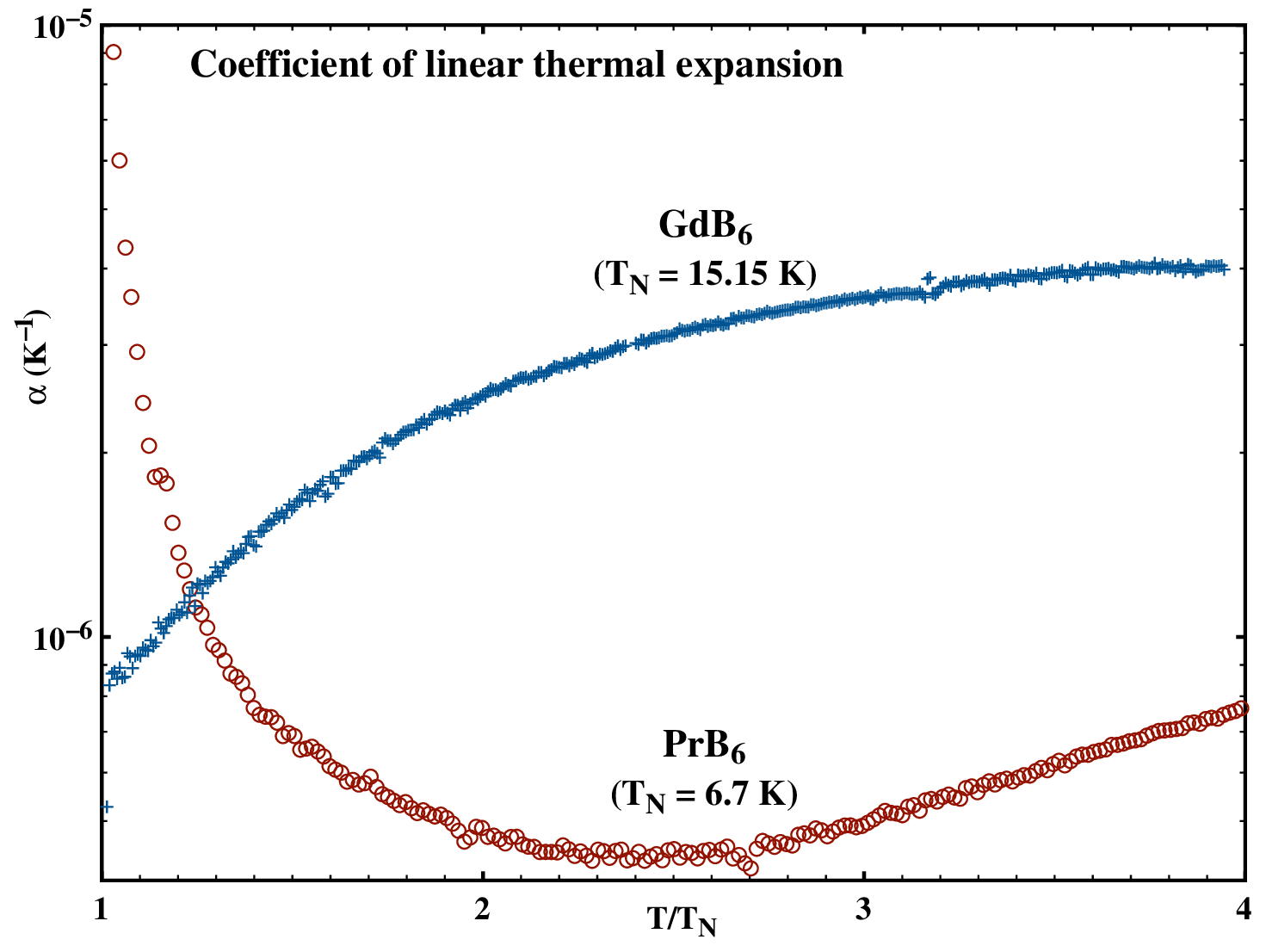}
\caption{\label{PrB6GdB6} Comparison of the linear thermal expansion coefficients $\alpha$ in the paramagnetic range of PrB$_6$ and GdB$_6$. The curves are obtained from measurements in equivalent conditions (cooling at 0.2 K/min) in the same setup, $\alpha$ being obtained from an unfiltered derivative of the relative length change as function of the temperature. $\alpha$ is plot over a logarithmic scale whereas the temperature is relative to the respective ordering temperatures T$_N$ (with values indicated between parentheses), in order to refer to the strength of the exchange interactions.}
\end{figure}

From neutron scattering~\cite{Loewenhaupt1986}, the Pr$^{3+}$ CEF ground state in PrB$_6$ has long been established as a well-isolated, $\Gamma_5$ magnetic triplet. However, previous experimental investigations, as well as the present one, reveal low-temperature paramagnetic behaviours that, below 30~K, are particularly difficult to reconcile with this simple scheme.
\begin{itemize}
\item[-] the first-order magnetic susceptibility is significantly reduced;
\item[-] the magnetic entropy decreases from the triplet value $R\ln3$ J/(K$\cdot$mol) at 30~K to about $R\ln2$ J/(K$\cdot$mol) just above $T_N = 6.7$~K;
\item[-] the thermal expansion coefficient increases dramatically on approaching $T_N$, where it becomes almost three orders of magnitude larger than in non-magnetic LaB$_6$.
\end{itemize}
In the presence of such paramagnetic anomalies, a natural explanation, close to a magnetic transition, is to invoke magnetic correlations. The continuous increase of the anomalies on approaching $T_N$ suggests precursor effects near a second-order magnetic transition, where the correlation length increases strongly. However, all experimental evidence points to a first-order antiferromagnetic transition in PrB$_6$, for which no critical divergence of the correlation length is expected. In PrB$_6$, no direct measurement of the correlation length is available in this temperature range. In the related compound Pr$_{0.8}$Ce$_{0.2}$B$_6$, a correlation length of approximately 15~\AA\ has been reported~\cite{Mignot2008} at 9.9~K, corresponding to only a few lattice spacings. This indicates that the magnetic correlations in this family remain relatively short-ranged, even close to the ordering temperature, whereas the observed anomalies in PrB$_6$ appear at temperatures exceeding 3$T_N$.

The thermal expansion anomaly is the most difficult to account for within the magnetic correlation hypothesis. In rare-earth compounds, magneto-elastic phenomena are generally discussed in terms of changes in the distribution of the 4$f$ shell electrons. For a rare-earth ion at a site of cubic symmetry, changes are essentially expected in the quadrupolar moments, which induce symmetry-breaking strains~\cite{MorinSchmitt1990}. These changes do not directly affect the volume, which couples to 4$^{\mathrm{th}}$- and 6$^{\mathrm{th}}$-order fully symmetric multipolar moments, those that develop under the influence of the cubic CEF. This causes a CEF thermal expansion contribution~\cite{Ott1977} that reflects the CEF level spacing, analogous to the Schottky contribution to the specific heat (see Fig.~\ref{Entropy}). In PrB$_6$, owing to the well-isolated CEF ground state, this effect should hardly be felt below~\cite{Novikov2017} 50~K.

In RE cubic compounds, the volume effects related to the magnetic ordering are small and are usually ascribed to exchange striction, i.e., to the dependence of the magnetic exchange interaction on the distance between rare-earth ions. The corresponding effect in the paramagnetic state, where only short-range magnetic correlations are present, should therefore be considerably smaller. To test this correlation hypothesis, one can turn to the element in the rare-earth hexaboride series where exchange striction is expected to be most pronounced, owing to the de Gennes rule and its higher transition temperature ($T_N = 15.2$~K): GdB$_6$. Fig.~\ref{PrB6GdB6} compares the thermal expansion coefficients, on a logarithmic scale, of PrB$_6$ and GdB$_6$, with the temperature normalised to their respective $T_N$ values. This is to emphasise the influence of the exchange coupling in the paramagnetic range, with the caveat that, in both cases, the magnetic transitions are first-order. One observes that, in GdB$_6$, where major correlation effects are expected, $\alpha$ decreases monotonically with temperature, resembling the behaviour observed in non-magnetic LaB$_6$ (Fig.~\ref{ThermExp}). This is in contrast to PrB$_6$, where $\alpha$ increases dramatically while approaching $T_N$, with a large anomaly exceeding one order of magnitude. It is therefore difficult to ascribe this large thermal-expansion anomaly in PrB$_6$ to magnetic correlations. Moreover, it simultaneously develops with the deviations in the magnetic susceptibility and magnetic entropy, which suggests a common origin. An alternative mechanism, one that does not rely on exchange couplings and related correlations, needs to be considered. In this regard, the absence of a thermal expansion anomaly in the orbital-singlet ($L=0$) compound GdB$_6$ favours a CEF-based interpretation for PrB$_6$ ($L=5$).

Indeed, all the anomalies discussed above could find a common explanation if one considers how the specific features of the cage environment change the CEF context. A distributed crystal-field model, as introduced in Ref.~\onlinecite{Amara2020} to describe similar anomalies in CeB$_6$, could naturally account for both the entropy loss and the reduction in magnetic susceptibility below 30 K. In CeB$_6$, these effects were reproduced by considering how the zero-point motion of the 4$f$ ion lifts the centre degeneracy of the non-Kramers CEF ground state. However, the model, which assumes a temperature-independent ground-state distribution of the cage oscillator, cannot account for a thermal-expansion anomaly. Within a single-site description, this would require the spatial distribution of the 4$f$ ion within the cage to evolve with temperature. This is precisely what one might qualitatively expect if, upon cooling, the lifting of the degeneracy at the cage centre leads to lower-energy states away from the centre.

In a forthcoming paper, in order to quantitatively test the cage crystal-field interpretation of the paramagnetic anomalies of PrB$_6$, the model developed for CeB$_6$ will be extended to account for the thermal-expansion effect.

\bibliography{/Applications/TeX/biblio}

\end{document}